\documentclass[%
 reprint,
superscriptaddress,
 amsmath,amssymb,
 aps,
]{revtex4-2}

	\usepackage{bm}
	\usepackage{braket}
	\usepackage{dsfont}
 	\usepackage{graphicx}
	\usepackage[caption=false]{subfig}
	\usepackage[export]{adjustbox}
	\newcommand{\vect}[1]{\boldsymbol{#1}}		% I use command \vect{} for vectors

\begin{document}
\title{Optical Signatures of F\"{o}rster-induced energy transfer in organic/TMD heterostructures}
\author{Joshua J. P. Thompson}
\email{joshua.thompson@physik.uni-marburg.de
}
\affiliation{Department of Physics, Philipps-Universit\"{a}t Marburg, Renthof 7, 35032 Marburg}
\author{Marina Gerhard}
\affiliation{Department of Physics, Philipps-Universit\"{a}t Marburg, Renthof 7, 35032 Marburg}
\author{Gregor Witte}
\affiliation{Department of Physics, Philipps-Universit\"{a}t Marburg, Renthof 7, 35032 Marburg}
\author{Ermin Malic}
\affiliation{Department of Physics, Philipps-Universit\"{a}t Marburg, Renthof 7, 35032 Marburg}
\begin{abstract}
    Hybrid van der Waals heterostructures of organic semiconductors and transition metal dichalcogenides (TMDs) are promising candidates for various optoelectronic devices, such as solar cells and biosensors. Energy-transfer processes in these materials are crucial for the efficiency of such devices, yet they are poorly understood. In this work, we develop a fully microscopic theory describing the effect of the F\"{o}rster interaction on  exciton dynamics and optics in a WSe$_2$/tetracene heterostack. We demonstrate that the differential absorption and time-resolved photoluminescence can be used to track the real-time evolution of excitons. We predict a strongly unidirectional energy transfer from the organic to the TMD layer. Furthermore, we explore the role temperature has in activating the F\"{o}rster  transfer and find a good agreement to previous experiments.  Our results provide a blueprint to tune the light-harvesting efficiency through temperature, molecular orientation and interlayer separation in TMD/organic heterostructures.      
\end{abstract}
\maketitle
The controlled fabrication of two-dimensional heterostructures has paved the way for new cutting-edge materials and quantum technologies \cite{geim2013van, liu2016van}. By combining materials with vastly different properties, novel device architectures can be designed to highlight and enhance desirable properties \cite{mueller2018exciton}. For typical, covalently bound monolayers, such as graphene, hBN and transition metal dichalcogenides (TMDs), the construction of van der Waals heterostructures has exciting physical and technological ramifications. From transistors \cite{wang2019van, leech2018negative} and superconductivity \cite{cao2018unconventional} to single photon emitters \cite{yu2017moire, baek2020highly} and sensors \cite{kim20202d}, these structures are at the forefront of scientific development. Furthermore, more exotic combinations of materials including TMD/organic \cite{bettis2017ultrafast, huang2018organic, zhu2018highly, thompson2023interlayer} and TMD/perovskite \cite{shi2018two, karpinska2022interlayer}  heterostructures, are areas of significant interest, not just in the 2D materials community but also as a way to enhance the properties of the partner material \cite{lin201917}.

Organic semiconductors (OSCs) have emerged as a leading candidate for a host of technological applications \cite{wilson2011ultrafast, congreve2013external}. Owing to their low-cost, scalable and environmentally-friendly fabrication  as well as their excellent light-harvesting properties, they are promising materials for photovoltaics \cite{darling2013case} and sensor applications \cite{lee2018recent}. The main drawback of these materials is their low efficiency, driven, in part, by their poor mobility \cite{moser2021challenges}. In a light-harvesting device, such as a solar cell, generated excitons (bound electron-hole pairs) can either recombine or separate into free charges at donor/acceptor interfaces or at the electrodes \cite{congreve2013external}. Charge separation is maximised when the mobility of the generated excitons is large allowing the charge carriers to be efficiently collected. 
One interesting approach to achieve this is to combine TMDs with OSCs \cite{sun20192d}.  The atomically flat interface of the TMDs acts as a good template on which to grow these crystals, and eliminates the effect of charge inhomogeneity \cite{huang2018organic}. Furthermore, depending on the band alignment of the chosen TMD and OSC, efficient charge and/or energy transfer can occur between the constituent layers. In a type-II heterostructure, charge transfer facilitates the formation of long-lived interlayer excitons \cite{jiang2021interlayer, zhu2018highly, thompson2023interlayer, ovesen2019interlayer, schmitt2022formation}.  Occurring in both type-I and -II heterostructures, energy transfer could allow excitons generated in the high absorption efficiency OSC to be transferred into the high mobility TMD \cite{kozawa2016evidence, gu2018dipole, ye2021ultrafast}. Both these schemes represent ways to use TMDs in order to enhance the light-harvesting efficiency of an OSC.  

In addition to light-harvesting, TMDs have been used as detectors of organic molecules. While this can be achieved by measuring a change in the electrical signal \cite{wang2022gate, kumar2022systematic}, significant performance has been achieved using optical methods \cite{feierabend2017proposal}. Organic molecules have been shown to modify the photoluminescence (PL) spectrum of a TMD \cite{feierabend2018molecule, ye2021ultrafast}. An exciting proposal is to use a layer of molecular aptasensors on the TMD as a selective probe of biological and infectious compounds \cite{zhou2020two, sen20222d}, such as those involved in malaria \cite{geldert2017based}, cancer \cite{kong2015novel}  and liver function \cite{shanmugaraj2019water}. In the absence of these compounds, energy transfer between the molecular layer and the neighbouring TMD quenches the signal of the aptasensor. At high enough concentration, the biological compound binds to the aptasensor triggering it to detach from the TMD surface and leading to a notable fluorescence/PL signal.  These devices represent selective probes of target biological molecules, which could be used as ultrathin, cheap, biological sensors, not only in the laboratory but also in the home and clinical settings. 

The fundamental mechanism describing the exciton energy transfer in these photovoltaic and biosensor systems is the F\"{o}rster interaction \cite{kozawa2016evidence, ye2021ultrafast}, with the energy transfer commonly described as F\"{o}rster-induced resonant energy transfer (FRET) \cite{zhou2020two}. Previous studies on two-dimensional systems found that the F\"{o}rster interaction dominates over Dexter-mediated charge transfer processes, in spite of the sub nm separation $d$. This can be attributed to the deviation from the $d^{-6}$ scaling law observed in localised molecular acceptor-donor systems, instead decaying more slowly at $e^{-d}$ for small distances and then $d^{-4}$ at larger distances. This was observed in graphene-TMD systems \cite{selig2019theory}, TMD-single molecule \cite{katzer2023impact} and molecule-graphene  \cite{malic2014forster, swathi2009distance}. Thus, in this study  we focus on the F\"{o}rster-mediated energy transfer. 
We develop a fully microscopic model describing the F\"{o}rster interaction between an exemplary OSC/TMD heterostructure comprised of WSe$_2$ and tetracene. Crucially, we describe the effect of the F\"{o}rster interaction on  optical spectra. We show pronounced F\"{o}rster-induced signatures in differential absorption and time-resolved PL spectra. We predict a strongly unidirectional energy transfer from the organic to the TMD layer. Furthermore, we find that the transfer rate drastically  increases with temperature due to the larger population of hot excitons.  Our results provide a recipe with which to enhance the F\"{o}rster interaction in a real heterostructure, through tuning the temperature, the interlayer distance, and the molecular orientation. By boosting the FRET in these heterostructures,  radiative losses e.g. in solar cells could be diminished, while the sensitivity of molecular aptasensors at different concentrations or orientations could be controlled. \\

\section{Results}
 \textbf{Tuning  of the F\"{o}rster transfer rate:}
 The F\"{o}rster interaction has already been intensively studied on a microscopic footing in quantum dots \cite{richter2006theory}, TMD-graphene \cite{selig2019theory} and molecule-graphene \cite{malic2014forster} systems, whereas the effect of the F\"{o}rster energy transfer has not been as well understood in the technologically promising TMD/OSC heterostructures.
A real space illustration of the F\"{o}rster interaction is shown in Fig 1(a) for a tetracene/WSe$_2$ heterostructure. The tetracene (Tc) in its bulk polymorph can be grown on top of a WSe$_2$ monolayer \cite{ye2021ultrafast, zhu2018highly, hummer2005electronic}. 

\begin{figure}[!t]
    \centering
\includegraphics[width=0.9\linewidth]{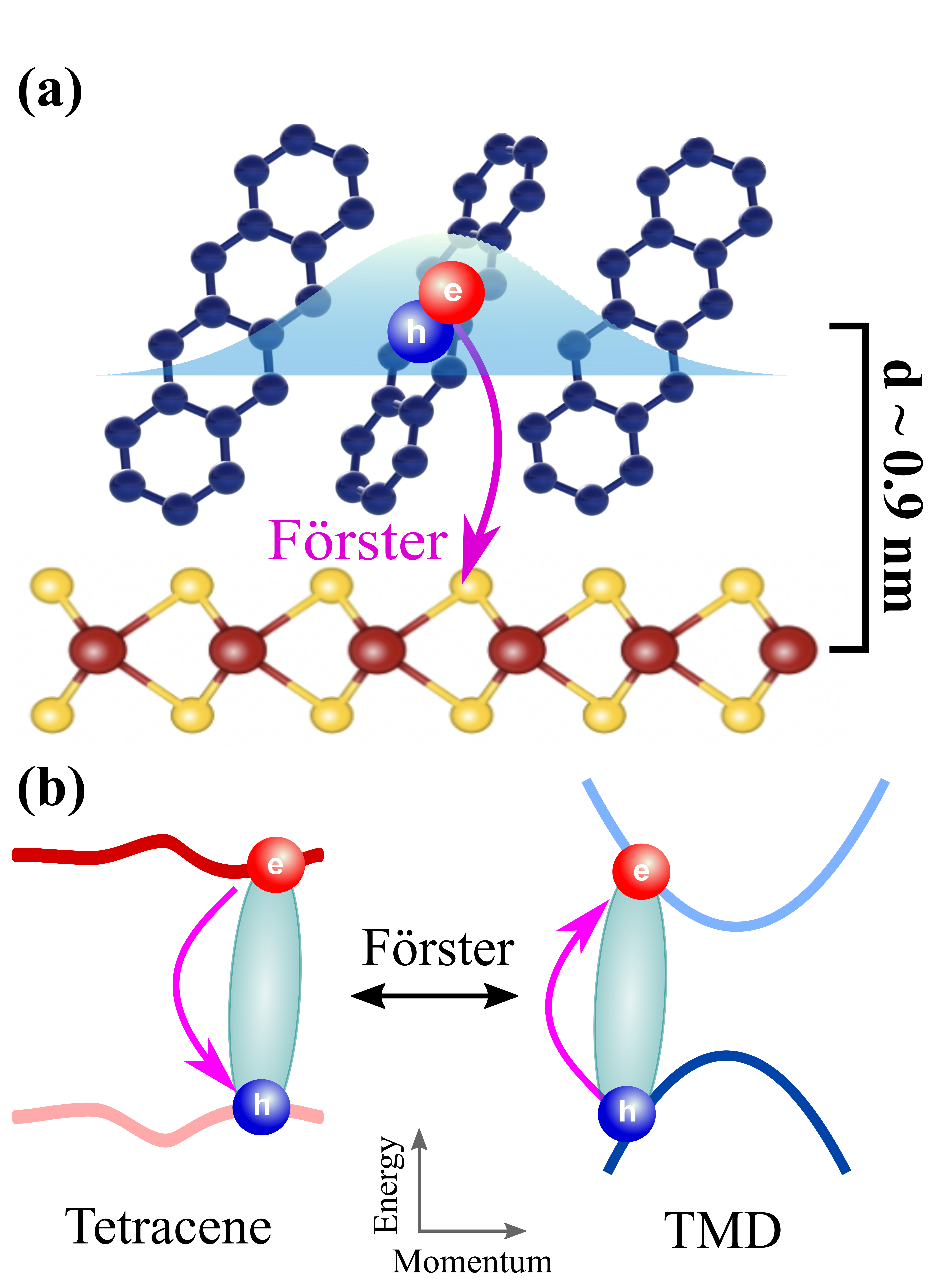}
    \caption{(a) Illustration of exciton energy transfer between the tetracene (blue) and the TMD layer.  (b) Single particle electronic band structure of tetracene (left) and WSe$_2$ (right) around the band gap. The F\"{o}rster interaction, conserving energy and centre-of-mass momentum, is shown by the pink arrows. }
    \label{fig:my_label}
\end{figure}

Using the semiconductor Bloch equations  and the Markov approximation \cite{kira2006many, brem2018exciton}, we can derive an expression for the exciton transfer rate due to the F\"{o}rster interaction \cite{selig2019theory}
\begin{align} \label{rate}
    \gamma_F = \dfrac{2}{\hbar}\dfrac{1}{Z}\sum_{\eta \vect{Q}} \Gamma^\eta_{\vect{Q}} \exp({-E^{ \text{Tc}}_{\vect{Q}, \eta}/k_BT}).
\end{align}
Here, $Z$ is the Boltzmann partition function, such that Eq. \ref{rate} is a thermal average of the F\"{o}rster dephasing rate $\Gamma^\eta_{\vect{Q}} = \pi \sum_{\mu } |V_{\mu \eta}(\vect{Q}, z)|^2 \delta(E^{ \text{Tc}}_{\vect{Q}, \eta} -  E^{ \text{WSe$_2$}}_{\vect{Q}, \mu})$ with $V_{\mu \eta}(\vect{Q}, z) = A^{\eta \mu}( \vect{Q}, z) \varphi^{\text{WSe$_2$}*}_{ \mu} (\vect{r}=0) \varphi^{\text{Tc}}_{, \eta}(\vect{r}=0) $ (c.f  Methods for more details).  This quantity describes exciton dephasing (and exciton linewidth broadening for $\vect{Q}=0$)  due to the interaction of an exciton in state $\eta$ in the Tc layer with momentum $\vect{Q}$ coupled to all states of the same momentum in state $\mu$ in the WSe$_2$ layer. The delta function ensures energy conservation, and coupled with the momentum and distance dependence contained in $V_{\mu \eta}(\vect{Q}, z)$ fully determines the exciton dephasing.
 The reverse process (WSe$_2$ to Tc) can be obtained by taking the thermal average over the TMD states.  The relatively flatter  bands in the Tc crystal lead to a broader momentum occupation and hence higher population of Tc excitons at larger $\vect{Q}$. As a result, the F\"{o}rster-mediated exciton transfer is much larger from the Tc to the WSe$_2$ layer than the other way round. We therefore predict a strong net transfer of excitons from the organic layer to the TMD.

\begin{figure}[!t]
    \centering
\includegraphics[width=\linewidth]{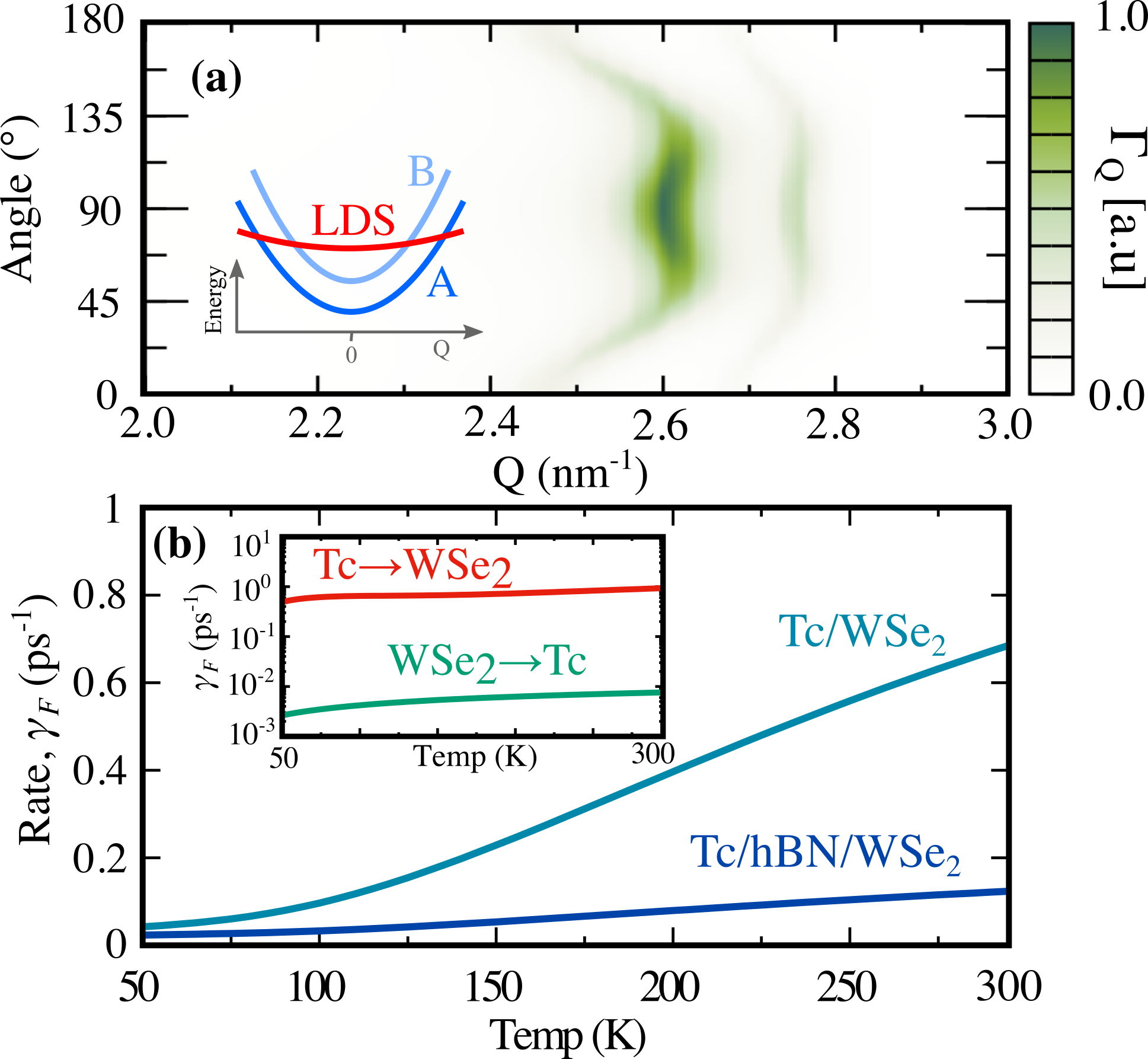}
    \caption{(a) F\"{o}rster-induced exciton dephasing $\Gamma_{\vect{Q}}$  in the lower Davydov state of the Tc crystal as a function of momentum direction (angle) and magnitude ($Q$). Inset shows schematic of excitonic band structures from the A (dark blue) and B (light blue) exciton in the TMD and the lower Davydov state (red) in the Tc layer. (b) F\"{o}rster-induced transfer rate from the Tc to the WSe$_2$ layer as a function of temperature for a direct Tc/WSe$_2$ interface and with an hBN spacer layer, respectively. The inset shows the temperature-dependent scattering rate both from the Tc to the WSe$_2$ layer (red) and from the WSe$_2$ to the Tc layer (green).}
    \label{fig:my_label}
\end{figure}
In Fig. 2 (a), the exciton dephasing of the lowest Davydov state is shown as a function of the center-of-mass momentum magnitude $Q$ and angle. The anisotropy of the Tc band structure gives rise to the characteristic shape of the intense region in (a), while the intensity varies in part due to the dipole orientation. The optical dipole of the lower Davydov state (LDS) is y-polarised (90$^\circ$), aligned along the $\vect{a}$-axis of the tetracene \cite{cocchi2018polarized}, and hence the strongest dephasing is observed in this orientation. This dipole dependence also explains the vanishing intensity at orthogonal angles (0$^\circ$ and 180$^\circ$). The two distinct bands appearing in Fig. 2 (a) originate from the coupling between the LDS in Tc and the A and B excitons in the TMD layer. The stronger emission stems from the coupling with the B exciton, which occurs at lower momenta as energy conservation occurs at lower $Q$ for this pair of bands. This can be seen in the inset of Fig 2(a), where the intersection of the excitonic dispersion (representing momentum and energy conservation) determines the shape of the features in Fig 2(a).  
In other many-particle processes, such as exciton-phonon scattering, the dephasing rate $\Gamma$  leads to a broadening of the exciton linewidth in PL and absorption spectra. In F\"{o}rster processes, however, this dephasing vanishes at $Q=0$ corresponding to the light-cone.  

Previous experimental studies have estimated the F\"{o}rster transfer rate, demonstrating a net transfer of excitons from a Tc layer into a WSe$_2$ layer \cite{ye2021ultrafast} with a transfer time of approximately 3 ps corresponding to a transfer rate of $\sim 0.29$ ps$^{-1}$. This occurs on a timescale much slower than typical exciton phonon-relaxation times in the material \cite{thompson2022singlet}, but also quicker than singlet fission and charge transfer \cite{ye2021ultrafast}. 
Evaluating Eq. (1), we calculate the  temperature dependence of the exciton transfer rate $\gamma_F$ from the Tc to the WSe$_2$ layer, cf.  Fig. 2 (b). We distinguish two cases: Tc directly on WSe$_2$ (light blue) and including an hBN spacer layer (dark blue).  As the temperature increases, the population of higher momentum excitons increases.  Since it is precisely these excitons that participate in the F\"{o}rster transfer process (note the linear Q dependence in Eq. (3)), the scattering rate also increases with temperature.  Adding additional hBN or organic spacer layers allows to tune the interlayer distance $z$.  We recover a $z^{-4}$ dependence  at large distances, characteristic of 2D systems \cite{malic2014forster, selig2019theory}, while at shorter distances a more rapid exponential decay is found.  As a result, we find  a lower F\"{o}rster rate ($\sim$ $5\times$ lower) in presence of a hBN spacer layer, cf. Fig 2. (b). This finding suggests that even in a multilayer Tc system, the main F\"{o}rster interaction occurs between the WSe$_2$ and the Tc layer at the interface. This and the fact that Tc excitons are strongly confined to individual Herringbone Tc layers \cite{cocchi2018polarized} suggests that our results can be extended to multi-layer Tc on WSe$_2$. \\

\textbf{F\"{o}rster visualized in differential absorption:}
The most convenient way to track the exciton transfer due to the F\"{o}rster interaction is by employing optical spectroscopy techniques.  The spectroscopy of choice should allow the population in one or both layers to be tracked. In differential absorption we can directly visualize the population change in either layer \cite{perea2020microscopic, katsch2019theory, selig2019ultrafast}.
The exciton absorption can be described using the well-established Elliot formula \cite{kira2006many, brem2018exciton}
\begin{align} \label{elliot}
    I_{\omega}(t) = \sum_\mu \dfrac{(1- B^\mu(t))|M_{\mu}|^2}{(E^{ \text{WSe$_2$}}_{0, \mu} - \hbar \omega)^2 +(\gamma_\mu+\Gamma^\text{phon}_\mu)^2}
\end{align}
where $\gamma_\mu$ and $\Gamma^\text{phon}_\mu$ are  the radiative and non-radiative (phonon) decay rates, respectively. The exciton-photon light-matter interaction $M_\mu$, is reduced by the bleaching term $
   B^\mu = \sum_{\vect{K}, \mu', \vect{q}} N^{\mu'}_{\vect{K}} (|\varphi^{\mu'}_{\vect{q}+\beta_{\mu'} \vect{K}}|^2 + |\varphi^{\mu'}_{\vect{q}-\alpha_{\mu'} \vect{K}}|^2) |\varphi^\mu(\vect{r}=0)|^{-1} \varphi^\mu_{\vect{q}}$ where $\alpha^\mu = m^\mu_c/M$ and  $\beta^\mu = m^\mu_v/M$.
This term crucially depends on the time-dependent excitonic population $N^{\mu'}_K$. 
We assume the optical excitation to be weak, such that the Coulomb renormalization of the excitonic resonance \cite{perea2020microscopic, katsch2019theory, selig2019ultrafast} is negligible and bleaching dominates. At much larger excitation fluences, clear resonance shifts should be observed due to population-induced screening and band renormalisation, while the bleaching term can become larger than 1, leading to optical gain.

\begin{figure} [!t]
    \centering
\includegraphics[width=0.95\linewidth]{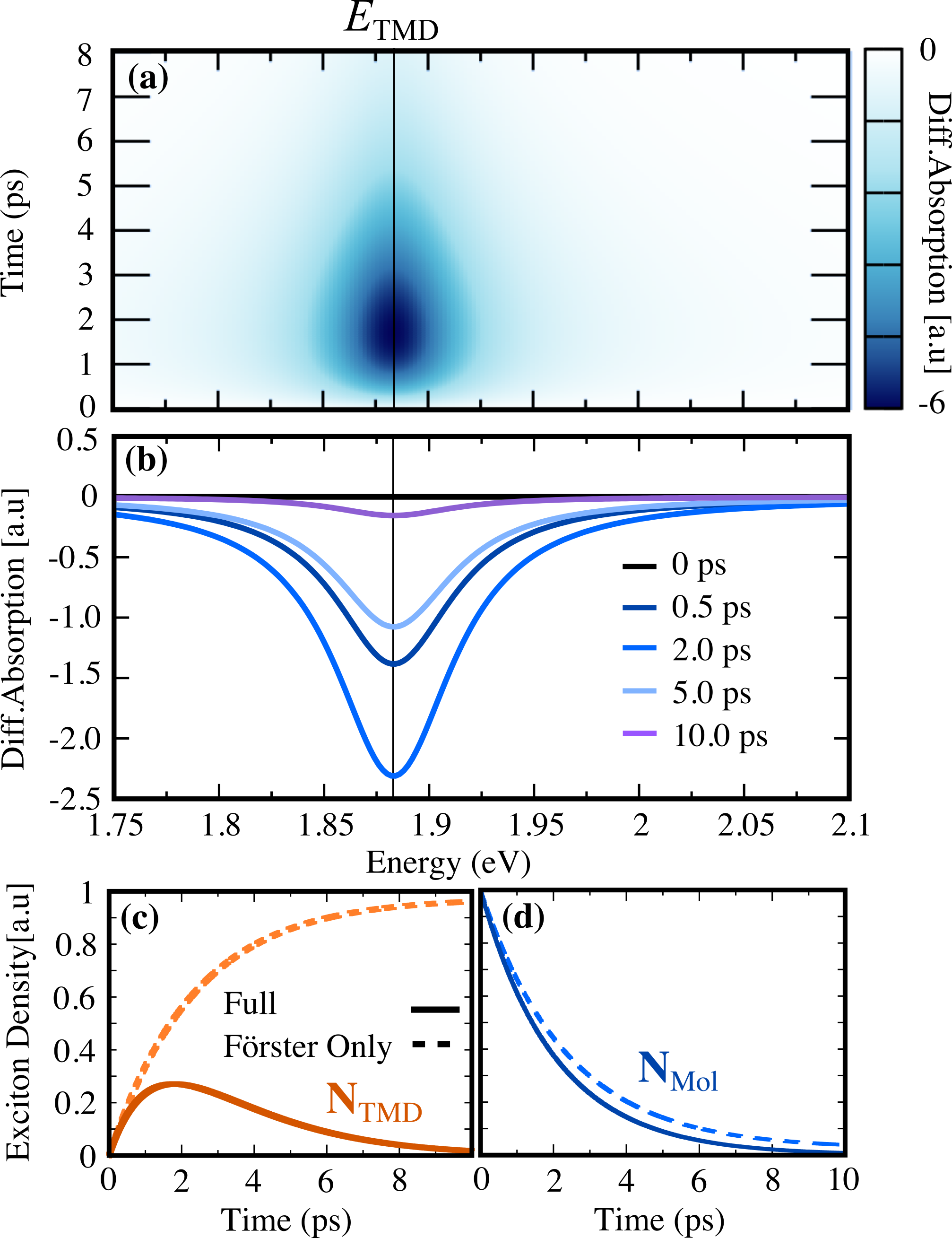}
    \caption{(a) Time-resolved differential absorption spectrum of the WSe$_2$ layer at 300 K. The lowest 1s exciton resonance is shown by the vertical line. (b) Time cuts of the differential absorption.  Temporal evolution of the exciton density in the (c) TMD and the (d) Tc layer, taking into account the full dynamics (solid) and without the radiative loss (dashed). }
    \label{fig:my_label}
\end{figure}

We derive a set of coupled semiconductor Bloch equations for the exciton polarization and incoherent exciton density, to describe this process (see SI for more details). We take explicitly into account in- and out-scattering between the incoherent excitons in the two layers due to both F\"{o}rster and exciton-phonon scattering. The excitonic properties, calculated using the Wannier equation, in addition to the phonon properties, allow us to determine not only the scattering dynamics but also the exciton broadening appearing in Eq. (\ref{elliot}).
The differential absorption $\Delta I_\omega(t) = I_\omega(t) -I_\omega(0)$ is defined as the difference in the absorption at time $t$ after excitation, and before the excitation $t=0$.
It is shown for a WSe$_2$/Tc heterostructure  in Fig 3(a). At time $t=0$, the differential absorption is 0 by definition. However, following an excitation pulse  resonantly exciting the Tc layer at $t=0$, the generated excitons transfer via the F\"{o}rster coupling to the WSe$_2$ layer. 
The phonon-driven intraband relaxation processes are, however, much faster than the F\"{o}rster interaction \cite{selig2019theory, thompson2022singlet}, such that  excitons in each layer remain thermalised. Using thermal averages, we find a simpler model for the exciton dynamics in the TMD ($l=0$) and Tc ($l=1$) layer
\begin{align} \label{dynam}
    &\partial_t N^{l} = -\gamma^{l}_\text{loss}
N^{l} -  \gamma_F^{l} N^{l}+ \gamma_F^{1-l} N^{1-l} 
\end{align}
such that in the bleaching term described above we define  $N^{\mu l}_{\vect{Q}} $ as $N^{\mu l}_{\vect{Q}}=\dfrac{N^{l}}{Z} \exp(-E^{ l}_{\vect{Q}, \mu}/k_BT)$.
The first term in Eq. (\ref{dynam})  describes the radiative and non-radiative loss \cite{brem2018exciton} scaling with the decay rate $\gamma_{\text{loss}}$. The second and third term describe the F\"{o}rster-induced out- and in-scattering,  respectively. These equations are accurate when the F\"{o}rster transfer occurs on a longer timescale than the phonon-scattering rate. 

In Fig. 3(a), we find a negative differential absorption of WSe$_2$ for times $t>0$, which is a  result of the exciton transfer from the Tc layer. Time cuts show a characteristic dip, cf. Fig 3(b). The differential absorption decreases between $t=0$ ps to around $t=2$ ps owing to the build up of excitons on the TMD layer. After around 2 ps, a steady state is reached between the exciton population in the WSe$_2$ and the one in the Tc layer, such that the decrease in the exciton population in both layers is driven primarily by the radiative and non-radiative decay. The temporal evolution of the exciton populations in the TMD  and molecular layer are shown in Fig 3 (c) and (d), respectively. We find that the population in the TMD layer initially increases before it drops, while the population of the Tc layer monotonously decreases. For the population in the TMD we find an interplay of F\"{o}rster induced charger transfer from the TC layer (population increases) and the radiative exciton recombination (population decreases). This results in a maximum population around 2 ps, cf. Fig. 3(c). In the absence of radiative recombination, we find a complete exciton transfer from the Tc to the WSe$_2$ layer (dashed lines). The pronounced dip in the differential absorption is a clear indication of the F\"{o}rster-induced energy transfer. The specific distance dependence of the F\"{o}rster interaction ($z^{-4}$ as explicitly shown in the SI) compared to other processes, such as tunneling or Dexter transfer \cite{malic2014forster}, allows us to identify the microscopic origin of the differential absorption signatures.\\

\textbf{F\"{o}rster visualized in time-resolved PL:} An additional probe of the F\"{o}rster-induced exciton dynamics can be achieved by using time-resolved photoluminescence  \cite{brem2018exciton, thompson2023interlayer}. As in the previous case, the LDS exciton in the Tc layer is resonantly excited leading to a thermalised exciton distribution on a short time-scale ($<$ 1 ps) \cite{thompson2022singlet}. The dynamics of the relative populations follow the results outlined in Fig 3 (c) and (d). The time-resolved PL is proportional to the exciton population within the light cone, $N^{\mu_l l}_{0}$, and is described using the  Elliot formula for PL \cite{brem2018exciton}
\begin{align}
 I^\text{PL}_{\omega}(t) = \sum_{\mu_l, l} \frac{N^{\mu_l l}_{0} |M^l_{\mu_l}|^2}{(E^{ \text{l}}_{0, \mu_l} - \hbar \omega)^2 +(\gamma_{\mu_l}+\Gamma^\text{phon}_{\mu_l})^2}  
\end{align}
 summing the contributions from both layers $l$. This equation differs from the aborption as it direcly depends on the exciton population $N^{\mu_l l}$.  The PL spectra are shown in Fig. 4(a) as a surface plot displaying the time and energy dependence. At time $t=0$, the exciton population is primarily in the Tc layer, so a clear signature of the lower Davydov state is observed.  After around 1 ps, the F\"{o}rster energy transfer occurs, depleting the signature from the LDS and leading to a pronounced increase in the PL of the A exciton in the WSe$_2$ layer.  Time snapshots are shown in Fig. 4(b). We estimate an energy transfer time of around 1-2 ps, which can be observed in the PL emission originating from the  WSe$_2$ reaching a maximum at around this time. We observe again how the PL signal begins to decrease after around 2 ps due to the exciton loss due to radiative and non-radiative decay. Differences in the decay rates will not change the qualtitative predictions and will only alter the time at which the strongest signal is observed, both in the PL and the differential absorption.

 \begin{figure} [!t]
    \centering
\includegraphics[width=0.95\linewidth]{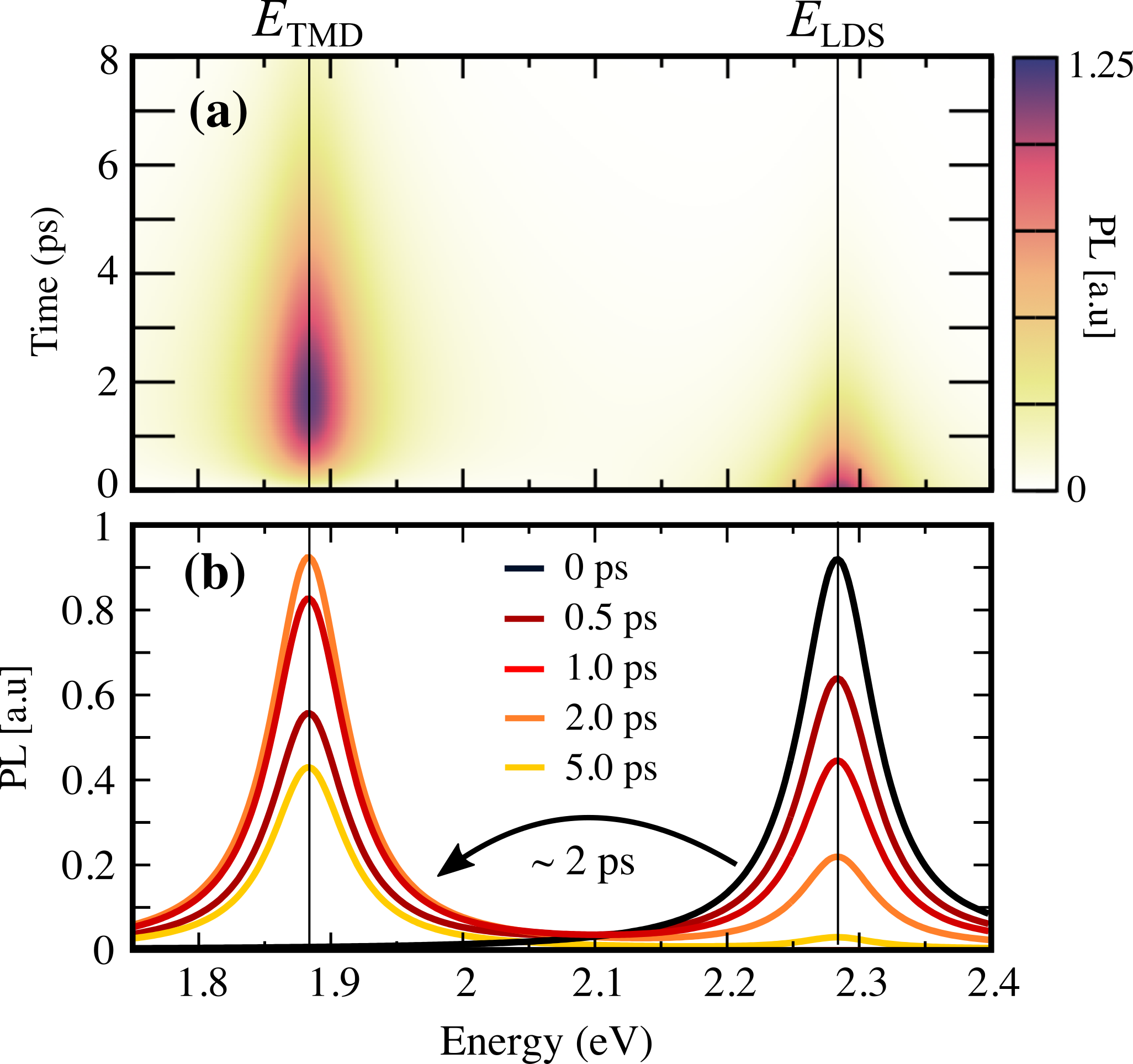}
    \caption{(a) Time- and energy-resolved photoluminescence spectrum of the WSe$_2$ and Tc layer at 300 K. The 1s exciton resonance in the TMD and the lower-Davydov state in the Tc layer are marked by vertical lines, respectively. (b) Snapshots of the PL at constant times, with an arrow marking the trend of exciton transfer due to the F\"{o}rster interaction. }
    \label{fig:my_label}
\end{figure}
\section{Discussion}
Depending on the structural geometry, namely the separation between the organic and TMD layers, different spectroscopic methods may be better suited to observe the F\"{o}rster-mediated energy transfer. When the interlayer separation is small, the rate is larger and higher-resolution methods, such as differential absorption, will be more appropriate. Similarly, in the case one or more spacer layers or lower temperatures, time-resolved PL may be better suited as higher time-resolution is less important.\\

 The F\"{o}rster coupling could be technologically exploited. The observed strong quenching of the PL signal from the molecular crystal is the crucial mechanism behind many optical biosensors \cite{zhou2020two, geldert2017based}. Here, in the absence of the target molecule,  the signature from the aptasensor adsorbed on the TMD is efficiently quenched. When a target molecule is present, the sensing molecules forming the molecular crystal, detach from the surface. This increases the molecular distance from the TMD, causing the F\"{o}rster coupling to quickly decrease, such that the molecular PL or fluorescence signal is no longer quenched. The coverage of molecules and the energetic/spatial separation between the molecular and TMD excitons allow the sensitivity of a biosensor to be tuned. Closer energetic alignment between exciton resonances will lead to even stronger exciton quenching. In our case, we used a well-known, exemplary organic crystal, however, our model can be easily extended to other organic molecules. Furthermore, an organic crystal is not strictly necessary, and similar results should be expected with a more sparse covering of molecules on the TMD.  The competition with temperature should also allow for a thermally-activated biosensor. Anisotropic materials, such as the Rhenium TMDs \cite{usman2022enhanced}, could also allow the orientation of the organic molecules relative to the TMD to be detected, due to the requirement that the TMD and organic dipoles should be aligned \cite{gu2018dipole}. In the same vein, the ''handedness" of chiral molecules could also be distinguished in the optical spectra \cite{zhao2019advances}, provided the F\"{o}rster interaction ocurrs on a shorter timescale than the intervalley relaxation in the TMD. \\
 
 In this work, we presented microscopic insights into the F\"{o}rster interaction between TMDs and organic crystals. Focusing on the exemplary WSe$_2$/Tc heterostructure, we show a characteristic  temperature and distance dependence of the F\"{o}rster-induced dephasing and exciton transfer. We estimate the energy transfer time and find a good agreement with previous experimental studies. The main message of this work lies in the question of how F\"{o}rster processes manifest in the optical spectra of an organic/TMD heterostructure. We demonstrate that the time-resolved differential absorption and photoluminescence  can capture well the F\"{o}rster-driven exciton dynamics, by tracking the change in the exciton population in the constituent layers.  Our work represents a significant advance in the microscopic understanding of organic/TMD interfaces and energy-transfer processes therein, important for fundamental physics but also for the design of new optoelectronic devices, such as biosensors.\\ 

 \section{Methods}
 \textbf{Microscopic Model for F\"{o}rster Transfer:} Following optical excitation, the generated electron-hole pairs interact via the screened Coulomb interaction forming excitons \cite{perea2022exciton}. We use the Wannier equation \cite{kira2006many} to calculate the exciton  energy dispersion $E^{ \text{WSe$_2$(Tc)}}_{\vect{Q}, \nu}$ and the corresponding excitonic wavefunctions $\varphi^{ \text{WSe$_2$(Tc)}}_{\vect{k}, \nu}$. The Wannier equation has been effective in determining the excitonic binding energy  in previous calculations on tetracene and pentacene, with a good agreement to experiments \cite{thompson2022singlet}. This is further justified by the fact that the excitonic wavefunction in oligoacenes, such as tetracene and pentacene, is dispersed over multiple neighbouring molecules and hence resembles a localised Wannier exciton \cite{sharifzadeh2018many, cocchi2018polarized, thompson2022singlet, alvertis2020impact}.

In tetracene, the symmetrically distinct molecules in the unit cell  gives rise to a Davydov splitting of the excitonic state \cite{cocchi2018polarized}. After photoexcitation, the exciton can then recombine, thermalise via scattering with phonons \cite{thompson2022singlet}, undergo the so-called singlet fission forming two long-lived triplet excitons \cite{thorsmolle2009morphology, congreve2013external} and in the  heterostructure case, take part in charge and energy transfer processes between the Tc and WSe$_2$ layers. Charge transfer, mediated by tunnelling/hopping processes, \cite{schmitt2022formation} gives rise to the formation of interlayer excitons. However,  this requires a significant wavefunction overlap, which is reduced in the considered herringbone structure, due to the relatively large interlayer separation of $0.9$ nm \cite{ye2021ultrafast}.   Furthermore, in the considered heterostructure Tc/WSe$_2$, the band alignment is such that charge transfer processes occur on a much slower time scale \cite{ye2021ultrafast} and hence the optically darker interlayer excitons are not expected to significantly contribute to optics nor the dynamics \cite{thompson2023interlayer}. 

On the other hand, the F\"{o}rster interaction depends less strongly on the interlayer separation \cite{malic2014forster, selig2019theory}, involving energy transfer between intralayer excitons in the constituent layers (cf. Fig 1(a)). In momentum space, the F\"{o}rster interaction involves the recombination (annihilation) of an exciton (electron and hole pair) in one layer and the creation of an exciton in other layer, mediated by the Coulomb interaction \cite{selig2019theory}. The Hamiltonian describing this process in the Tc/WSe$_2$ heterostructure can be written as, 
\begin{align}
  H_{F} = \sum_{\substack{\lambda,\lambda',\nu,\nu'\\ \vect{k}, \vect{k'}, \vect{q}, \vect{q'}}} V^{\vect{k}, \vect{k'}, \vect{q}, \vect{q'}}_{\lambda,\lambda',\nu,\nu'} \hat{a}^\dagger_{\lambda, \vect{k}} \hat{B}^\dagger_{\nu, \vect{q}} \hat{B}_{\nu', \vect{q'}} \hat{a}_{\lambda', \vect{k'}} +h.c,
\end{align}
where $\hat{a}^{(\dagger)}$ and $\hat{B}^{(\dagger)}$ are annihilation (creation) operators acting on the Tc and WSe$_2$ layer, respectively. The indices $\lambda$ and $\nu$ describe the hybrid spin/valley/band index. The Coulomb matrix element $V^{\vect{k}, \vect{k'}, \vect{q}, \vect{q'}}_{\lambda,\lambda',\nu,\nu'}$ is screened by the surrounding dielectric, $\epsilon$, of a  SiO$_2$ substrate \cite{ye2021ultrafast}. 
Using a real-space coordinate transformation  the Coulomb matrix element is 
$V^{\vect{k}, \vect{k'}, \vect{q}, \vect{q'}}_{\lambda,\lambda',\nu,\nu'}= \sum_{\vect{s} \vect{Q}} e^{-i \vect{Q} (\vect{s}+\vect{z})}\delta_{\vect{Q}, \vect{k}-\vect{k'}} \delta_{\vect{Q}, \vect{q'}-\vect{q}} I_{\vect{k}, \vect{q}, \vect{Q}}^{\vect{s}, \vect{z}} 
 $,     where $ I_{\vect{k}, \vect{q}, \vect{Q}}^{\vect{s}, \vect{z}}$ takes the form of a dipole-dipole interaction (cf. the SI), separated into in-plane, $\vect{s}$, and out-of-plane, $\vect{z}$, components.
 
  Since the dominant F\"{o}rster transfer processes occur between the lowest lying energy levels we consider only the highest valence/lowest conduction band of the WSe$_2$ and the HOMO/LUMO of the Tc.  We only consider spin-like states,  as the Coulomb matrix elements  vanish for excitons comprised of opposite spins \cite{wang2017plane}. The same applies to momentum dark intervalley excitons in both layers \cite{perea2022exciton}.
In the tetracene layer, the upper and lower Davydov excitons originate from the intermolecular coupling between the symmetrically distinct molecules in the unit cell. The resulting lower and upper Davydov excitons form around the M and $\Gamma$ point respectively \cite{cocchi2018polarized}. 
Taking this into account, we arrive at the F\"{o}rster Hamilton operator  in the excitonic basis (see the SI for a detailed derivation)
\begin{align}
    H_{F} =   \sum_{\eta, \mu, \vect{Q}} A^{\eta \mu}( \vect{Q}, z) \varphi^{\text{WSe$_2$}*}_{ \mu} \varphi^{\text{Tc}}_{\eta} \hat{X}^{\text{WSe$_2$}\dagger}_{\vect{Q}\mu} \hat{X}^{\text{Tc}}_{\vect{Q}\eta} + h.c.
\end{align}
 where $\hat{X}^{\text{l}(\dagger)}_{\vect{Q}\mu}$ are exciton annihilation (creation) operators with momentum $\vect{Q}$, and energy level $\mu$, on layer $l$. Here, we can express the excitonic wavefunction in real-space coordinates as $\sqrt{A}\varphi_{\mu/\eta} (\vect{r}=0)  =\sum_{\vect{k}}\varphi_{\vect{k}, \mu/\eta} $, were the hybrid indices $\lambda$, $\nu$ have been absorbed in the excitonic indices $\mu$ and $\eta$.
The appearing matrix element $A^{\eta \mu}( \vect{Q}, z) $ can be expressed  analytically (cf. the SI).
\begin{align} 
   A^{\eta\mu}(\vect{Q}, z)= \frac{e^{-i Q_z z} e^{-|z|Q}}{4\epsilon_0 \epsilon} \vect{d}^{\text{WSe$_2$}}\cdot \vect{Q}\left(\frac{\vect{Q}}{|\vect{Q}|}+i\hat{z}\right)\cdot\vect{d}^{\text{Tc}}.
\end{align}
Here, we have introduced the optical transition dipoles $|\vect{d}^{\text{WSe$_2$}}| = 0.4$ nm \cite{selig2019theory} and $|\vect{d}^{\text{Tc}}| = 0.05$ \cite{pati2014exact} and are assumed to be approximately constant around the WSe$_2$ and Tc valleys, respectively \cite{selig2019theory}. The out-of-plane unit vector $\hat{z}$ and out-of plane momentum $Q_z$ also appear in the the dipole-dipole interaction, and become more relevant in structures with organic dipoles pointing perpendicularly to the layer plane.

Encoded in this Hamiltonian is the conservation of the center-of-mass momentum between the excitons in the WSe$_2$ and Tc. This is shown schematically in Fig 1(b). Importantly, the matrix element $A(\vect{Q}, z)$  describes the dependence on the exciton momentum and interlayer separation $z$.  We find that the F\"{o}rster interaction vanishes as $\vect{Q} \rightarrow 0$, which is typical for dipole-dipole-type coupling.  The transition dipole of the TMD is circularly polarised, with handedness depending on the valley, which give rise to the well-known optical valley selection rules in TMDs \cite{zeng2012valley}. In contrast, the lowest Davydov excitons in Tc are linearly polarised \cite{pati2014exact, camposeo2010polarized}. As a result the F\"{o}rster matrix element becomes anisotropic in $\vect{Q}$, favouring the transition dipole of the initial or final Davydov exciton. Unless otherwise stated, we take a typical interlayer separation of 0.9 nm, defined as the distance between the W-atom layer in the middle of the  WSe$_2$, and the centre of Tc layer (c.f. Fig 1 (a)).\\

\noindent \textbf{Acknowledgements}   This project has received funding from Deutsche Forschungsgemeinschaft via CRC 1083 and the European Unions Horizon 2020 research and innovation programme under grant agreement no. 881603 (Graphene Flagship).
 \noindent \textbf{Data Availability} 
  The data that support the findings of this study are available from the corresponding author upon reasonable request.

  \noindent \textbf{Competing Interests}
  The authors declare no competing interests.
  
\noindent \textbf{Author Contributions} E.M conceived the project. J.J.P.T developed the theory and performed the calculations. E.M, J.J.P.T, M.G and G.W analyzed the obtained results. J.J.P.T wrote the paper with all the authors contributing to the discussion and preparation of the manuscript. 

\bibliographystyle{achemso}
\providecommand{\latin}[1]{#1}
\makeatletter
\providecommand{\doi}
  {\begingroup\let\do\@makeother\dospecials
  \catcode`\{=1 \catcode`\}=2 \doi@aux}
\providecommand{\doi@aux}[1]{\endgroup\texttt{#1}}
\makeatother
\providecommand*\mcitethebibliography{\thebibliography}
\csname @ifundefined\endcsname{endmcitethebibliography}
  {\let\endmcitethebibliography\endthebibliography}{}

\end{document}